\documentclass[manuscript]{aastex}
\usepackage{amssymb}
\usepackage{txfonts}

\usepackage{amsmath}
\usepackage{float}

\shortauthors{S. Liu et al.}

\begin{document}

\title{A Statistical Study on Force-Freeness of Solar Magnetic Fields in the Photosphere}

\author{S. Liu\altaffilmark{1}, J.T. Su\altaffilmark{1}, H.Q. Zhang\altaffilmark{1},
 Y.Y. Deng\altaffilmark{1}, Y. Gao\altaffilmark{1}, X. Yang\altaffilmark{1}, X.J. Mao\altaffilmark{1,2}}

\affil{$^{1}$National Astronomical Observatory and key Laboratory of Solar Activity, \\Chinese Academy of
Sciences,
        Beijing 100012, China}

\affil{$^{2}$Department of Astronomy, Beijing Normal University, Beijing 100875, China}

\email{lius@nao.cas.cn}

\altaffiltext{}{key Laboratory of Solar Activity}

\begin{abstract}
It is an indisputable fact that solar magnetic fields are force-free
in the corona, where force free fields means that current and magnetic
fields are parallel and there is no Lorentz force in the fields. While
the force-free extent of photospheric magnetic fields remains open.
In this paper, the statistical results about it
is given. The vector magnetograms (namely, $B_{x}$, $B_{y}$ and $B_{z}$ in heliocentric coordinates) are employed,
which are deduced and calibrated from Stokes spectra, observed by Solar
Magnetic Field Telescope (SMFT) at Huairou Solar Observing Station
(HSOS) are used. We study and calibrated 925 magnetograms calibrated by two
sets of calibration coefficients, that indicate the relations between magnetic fields
and the strength of Stokes spectrum and can be calculated either theoretically or empirically.
The statistical results show that the majority of active region magnetic fields are not
consistent with the force-free model.
\end{abstract}


\section{Introduction}

Magnetic fields dominate most solar activites such as filaments eruption, flares and coronal
mass ejections (CMEs). All these phenomena are energetic events due to explosive release of magnetic
energy \citep{kra82, shi95, tsu96, wan94, bon01, pri03, nin12}. Thus the solar magnetic field
is the key for us to understand the nature of solar activities.
Since the measurable magnetic sensitive lines are around the photosphere,
the reliable measurements of magnetic fields are nearly there \citep{ste73, har77}.
However the understanding of the magnetic field in the chromosphere and corona remains difficult due to
both intrinsic physical difficulties and observational limitations \citep{gar94, lin04}.
Generally, magnetic fields in the corona are regarded as a force-free
\citep{aly89}, because the plasma $\beta$ (ratio of plasma pressure
to magnetic pressure over there) is much less than unity. But it is controversial in the photosphere,
because two kinds of pressure are comparable \citep{dem97}. For the low-$\beta$ corona where the plasma
is tenuous ($\beta \ll 1$), magnetic field satisfies the following
force-free equations:
\begin{equation}
\nabla \times \textbf{B} = \alpha(\textbf{r}) \textbf{B},
\end{equation}
\begin{equation}
\nabla \cdot \textbf{B} = 0.
\end{equation}
They imply that there is no Lorentz force in action and $\alpha$ is constant
along magnetic field lines (\textbf{B}$\cdot\nabla\alpha = 0$).

At present, the magnetic field extrapolation with a force-free
assumption is a major method to study the solar magnetic
fields of active regions. The coronal fields can be reconstructed from a physical
model (namely, force free model) in which the observed photospheric magnetic field is taken
as a boundary condition \citep{wu90, mic94, ama97, sak81, yan00, whe00, wie04, son06, he08, liu11a}.
 It means that coronal magnetic fields are considered to be force-free, while at the boundary it is
 connected to the photospheric magnetic fields observed. The
force-free extent of the photospheric magnetic field then becomes an important
subject to study. \citet{wie06} proposed a preprocessing
procedure to make a minor regulation within the allowable errors, so that the observed magnetic fields are biased to
a force-free, which further indicates
the study of force-free extent is significant and necessary for the field extrapolation. \citet{mat95}
calculated dependence of the net Lorentz force in the
photosphere and low chromosphere on the height using Mees Solar Observatory
magnetograms and concluded that the magnetic fields are not force
free in the photosphere, but becomes force-free roughly 400 km above
the photosphere. \citet{moo02} studying the force-free extent
in the photosphere using 12 vector magnetograms of three active regions,
realized that the photospheric magnetic fields are not very
far from the force-free case. \citet{liu11b} tentatively applied the force-free extrapolation to
reconstruct the magnetic fields above the quiet region  and checked the force-free extent of this
quiet region based on the high spatial resolution vector magnetograms observed by the Solar
Optical Telescope/Spectro-Polarimeter on board Hinode. \citet{tiw12}
found sunspot magnetic fields are not so far from the force-free.
We, in this paper, conduct a the statistical research to make use of the vector
magnetograms observed by SMFT at HSOS from 1988 to 2001 in
order to verify the force-freeness of the photospheric magnetic field.

The paper is organized as follows: firstly, the description of
observations and data reduction is arranged in Section~\ref{S-Obser and Data};
then, the results are shown in Section~\ref{S-Results};
finally, in section~\ref{S-Conl} presents the short discussions and
conclusions.

\section{Observations and Data Reduction}
\label{S-Obser and Data}

The observational data used here are 925 vector magnetograms corresponding to 925
active regions observed from 1988 to 2001 by olar
Magnetic Field Telescope (SMFT) installed at Huairou Solar Observing Station (HSOS),
located at north shore of Huairou reservoir. Magnetograms associated with the corresponding to
the active region, which is the nearest to disk
center is chosen and calculated. So only one magnetogram is available for one active region. SMFT consists of 35 cm refractor
with a vacuum tube, birefringent filter, CCD camera including an image processing system
operated by computer \citep{ai86}. The birefringent filter is tunable, working either at the photosphere line Fe I $\lambda$ 5324.19 \AA, with a 0.150 \AA ~bandpass, or at the chromosphere line, H$\beta$, with a 0.125 \AA ~bandpass. The line of Fe I $\lambda$ 5324.19 \AA ~(Lande factor $g$
=1.5) formed around the solar photosphere, is used for photospheric magnetic field observations. The
bandpass of the birefringent filter is about 0.15 \AA~for Fe I $\lambda$~5324.19
\AA~line. The center wavelength of the filter can normally be shifted -0.075 \AA~ relative to center of Fe I $\lambda$~
5324.19 \AA~ to measure of the longitudinal magnetic field and then the line center is applied to measure the transverse one \citep{ai86}.
Vector magnetograms are reconstructed from four narrow-band images of Stokes parameters ($I$, $Q$, $U$
and $V$). $V$ is the difference of the left and right circularly polarized images, $Q$ and $U$ are the differences between
two orthogonal linearly polarized images for different azimuthal directions, $I$ is the intensity derived from either the sum of two circularly polarized images is the line-of-sight field measurements or of two linearly polarized images in the transverse field measurement.
When $I$, $Q$, $U$ and $V$ are measured, the corresponding white light images are simultaneously obtained, which are employed to compensate for
the time differences during the measurements of $I$, $Q$, $U$ and $V$.
The sequence of obtaining Stokes images is as follows: First acquired the V/I image
; next the Q/I images, then the U/I image. The time required to
obtain a set of Stokes images was about 45 second. Each
image is associated with 256 integrated frames. To reconstruct the vector magnetograms, the linear relation is necessary between the magnetic field and the Stokes parameters $I$, $Q$, $U$
and $V$, which is true under the weak-field approximation \citep{jef89, jef91}:
\begin{equation}
B_{L}=C_{L}V ,\\
\end{equation}
\begin{equation}
B_{T}=C_{T}(Q^{2}+U^{2})^{1/4}, \\
\end{equation}
\begin{equation}
\theta=arctan(\dfrac{B_{L}}{B_{\bot}}),\\
\end{equation}
\begin{equation}
\phi=\dfrac{1}{2}arctan(\dfrac{U}{Q}),\\
\end{equation}
where $B_{L}$ and $B_{T}$ are the line-of-sight and transverse component of the photospheric field, respectively.
$\theta$ is the inclination between the vector magnetic field and the direction normal to the solar surface and $\phi$ is the field azimuth.
$C_{L}$ and $C_{T}$ are the calibration coefficients for the
longitudinal and transverse magnetic fields, respectively. Both theoretical and empirical
methods are used to calibrate vector magnetograms \citep{wan96, ai82}. So that two sets of calibration coefficients are available:
the first set $C_{L}$ and $C_{T}$ are 8381 G and 6790 G \citep{su04,wan96},
respectively, obtained by theoretical calibration; in the
second set $C_{L}$ and $C_{T}$ are 10000 G and 9730 G, respectively,
which are deduced through empirical calibrations \citep{wangjx96}.
Faraday rotation and magneto optical effects may affect the value of
measured magnetic fields. \citet{bao00} analyzed the Faraday rotation in the
HSOS magnetograms which contributes about $12^{\circ}$. \citet{gao08}
proposed a way of the statistical removal of Faraday rotation in vector
magnetograms from HSOS. \citet{zhang00} found that the
statistical mean azimuth error of the transverse field is $12.8^{\circ}$ caused
by magneto optical effects. In addition, there may be
other causes of the data uncertainty. For
example, saturation effects, filling factors and stray light. After routine data processing of HSOS data, the spatial
resolution of observational data is actually 2 arcsec/pixel $\times$
2 arcse/pixel and 3$\sigma$ noise levels of vector magnetograms are 20
G and 150 G for longitudinal and transverse components,
respectively.

The acute angle method is employed to
resolve $180^{\circ}$ ambiguity \citep{wang94,wang97,wang01,mat06}, in which the observed field is
compared to the extrapolated potential field in the photosphere. The
orientation of the observed transverse component is chosen by
requiring $-90^{\circ}$ $\leq$ $\bigtriangleup \theta$ $\leq$
$90^{\circ}$, where $\bigtriangleup \theta$ =
$\theta_{o}$-$\theta_{e}$ is the angle difference between the observed and
extrapolated transverse components. At last, the components of the vector magnetic field
$(B_{x}=B_{T}cos\phi$, $B_{y}=B_{T}sin\phi$ ($B_{T}=\sqrt{B_{x}^{2}+B_{y}^{2}}$) and $B_{z}=B_{L}cos\theta$) are calculated in local heliocentric coordinates. To minimize the projection
effects, the requirement that the horizontal width of an active region is less
than 20 degree is added for each magnetogram.

\section{Results}
\label{S-Results}

In classical electromagnetic theory, Lorentz force can be written as the divergence of the Maxwell stress:
\begin{equation}
\label{ff} \textbf{F} = \dfrac{(\textbf{B}\cdot \nabla)\textbf{B}}{4\pi}-\dfrac{\nabla (\textbf{B}\cdot \textbf{B})}{8\pi},
\end{equation}
Under the assumption that the magnetic field
above the plane z = 0 (namely, the photosphere) falls off fast enough as going
to infinity, the net Lorentz force in the infinite half-space z $>$ 0 is
just the Maxwell stress integrated over the plane z = 0 \citep{cha61, mol74, aly84, low85}. Then, the components of the net Lorentz force at the
plane z = 0 can be expressed by the surface integrals as follows:
\begin{equation}
\label{fxyzp01} F_{x} = -\dfrac{1}{4\pi}\int B_{x}B_{z}dxdy,
\end{equation}
\begin{equation}
\label{fxyzp02}F_{y} = -\dfrac{1}{4\pi}\int B_{y}B_{z}dxdy,
\end{equation}
\begin{equation}
\label{fxyzp03}F_{z} = -\dfrac{1}{8\pi}\int (B_{z}^{2}-B_{x}^{2}-B_{y}^{2})dxdy.
\end{equation}

The necessary conditions of a force-free field are that all three components are much less than $F_{p}$ \citep{low85}, where $F_{p}$ is a characteristic magnitude of the total Lorentz force and can be written as:
\begin{equation}
\label{fxyzp1} F_{p} = \dfrac{1}{8\pi}\int
(B_{z}^{2}+B_{x}^{2}+B_{y}^{2})dxdy,
\end{equation}
The values of $F_{x}/ F_{p}$, $F_{y}/ F_{p}$, and $F_{z}/ F_{p}$ then provide a measure of the force-free extent at the boundary plane z = 0
(the photosphere).

Following \citet{mat95} and \citet{moo02}, $F_{x}/ F_{p}$, $F_{y}/ F_{p}$, and $F_{z}/ F_{p}$ are utilized to check the force-free extent of the selected
photospheric magnetograms. The necessary
conditions of the force-free field are also expressed as:
$|F_{x}|/ F_{p}\ll 1$, $|F_{y}|/ F_{p} \ll 1$, and $|F_{z}|/ F_{p}\ll 1$
\citep{mat95,moo02}, that is if three parameters
$F_{x}/F_{p}$, $F_{y}/F_{p}$ and $F_{z}/F_{p}$ are so small that they are negligible then the
magnetic field can be regarded as a force-free completely. \citet{mat95} suggested that the magnetic field is a force-free
if the $F_{z}/F_{p}$ is less than or equal to
0.1. The calibration coefficients $C_{L}$ and $C_{T}$ may affect the three parameters of $F_{x}/F_{p}$, $F_{y}/F_{p}$ and
$F_{z}/F_{p}$. Thus, two sets of calibration
coefficients mentioned above are applied to this study. CaseI: $C_{L}$ and $C_{T}$ are chosen as 8381 G and
6790 G \citep{su04,wan96}, respectively. CaseII: $C_{L}$ and $C_{T}$
are chosen as 10000 G and 9730 G \citep{wangjx96}, respectively.

\begin{figure}

   \centerline{\includegraphics[width=1.\textwidth,clip=]{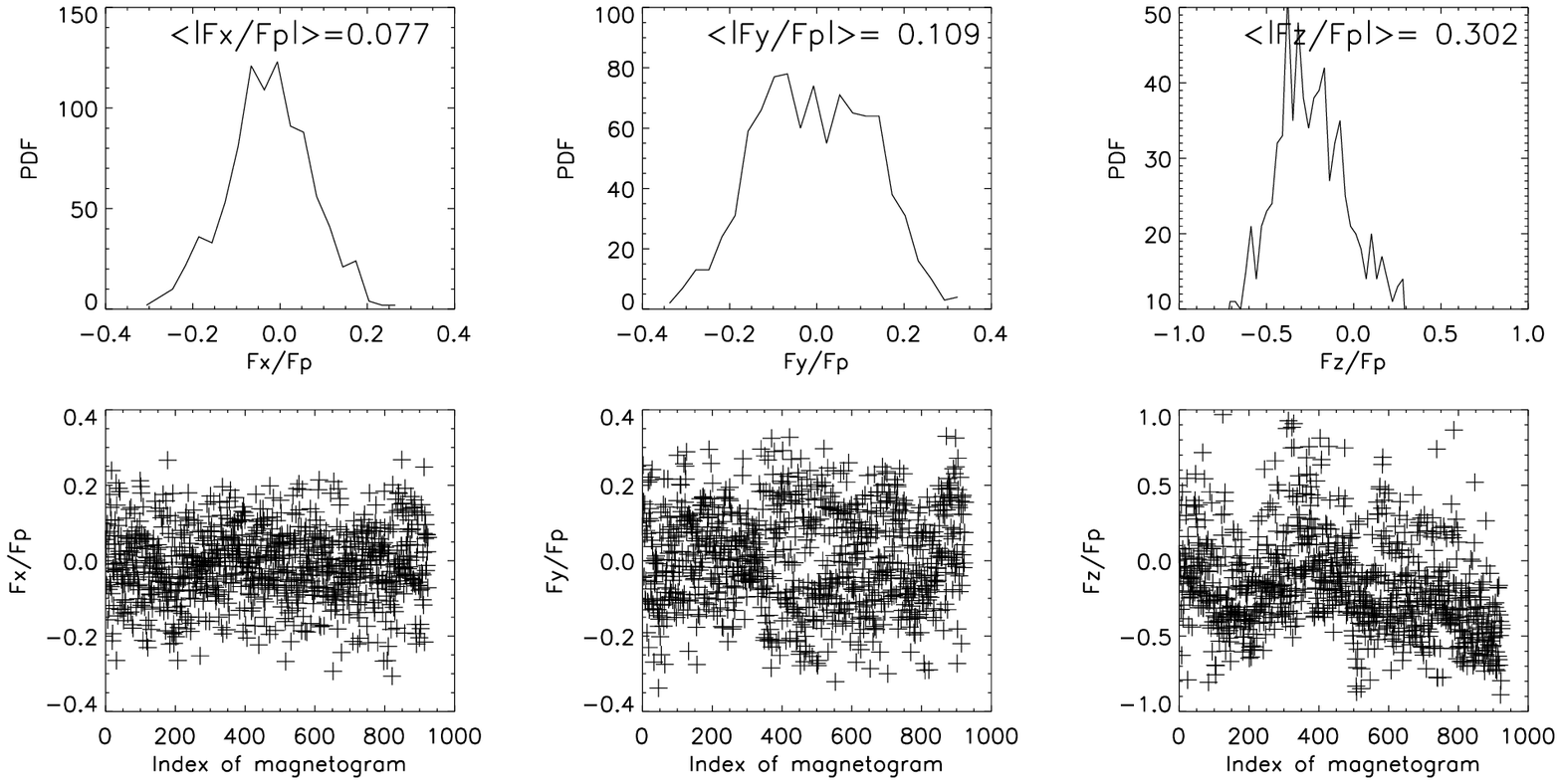}}

   \caption{PDF and scatter diagrams of $F_{x}/F_{p}$, $F_{y}/F_{p}$ and $F_{z}/F_{p}$ for the selected
   magnetograms. The mean values of absolute $F_{x}/F_{p}$, $F_{y}/F_{p}$ and
   $F_{z}/F_{p}$ are plotted and indicated by $<|F_{x}/F_{p}|>$,$<|F_{y}/F_{p}|>$, $<|F_{z}/F_{p}|>$, respectively. (CaseI)
   } \label{Fig1}
\end{figure}
\begin{figure}

   \centerline{\includegraphics[width=1.\textwidth,clip=]{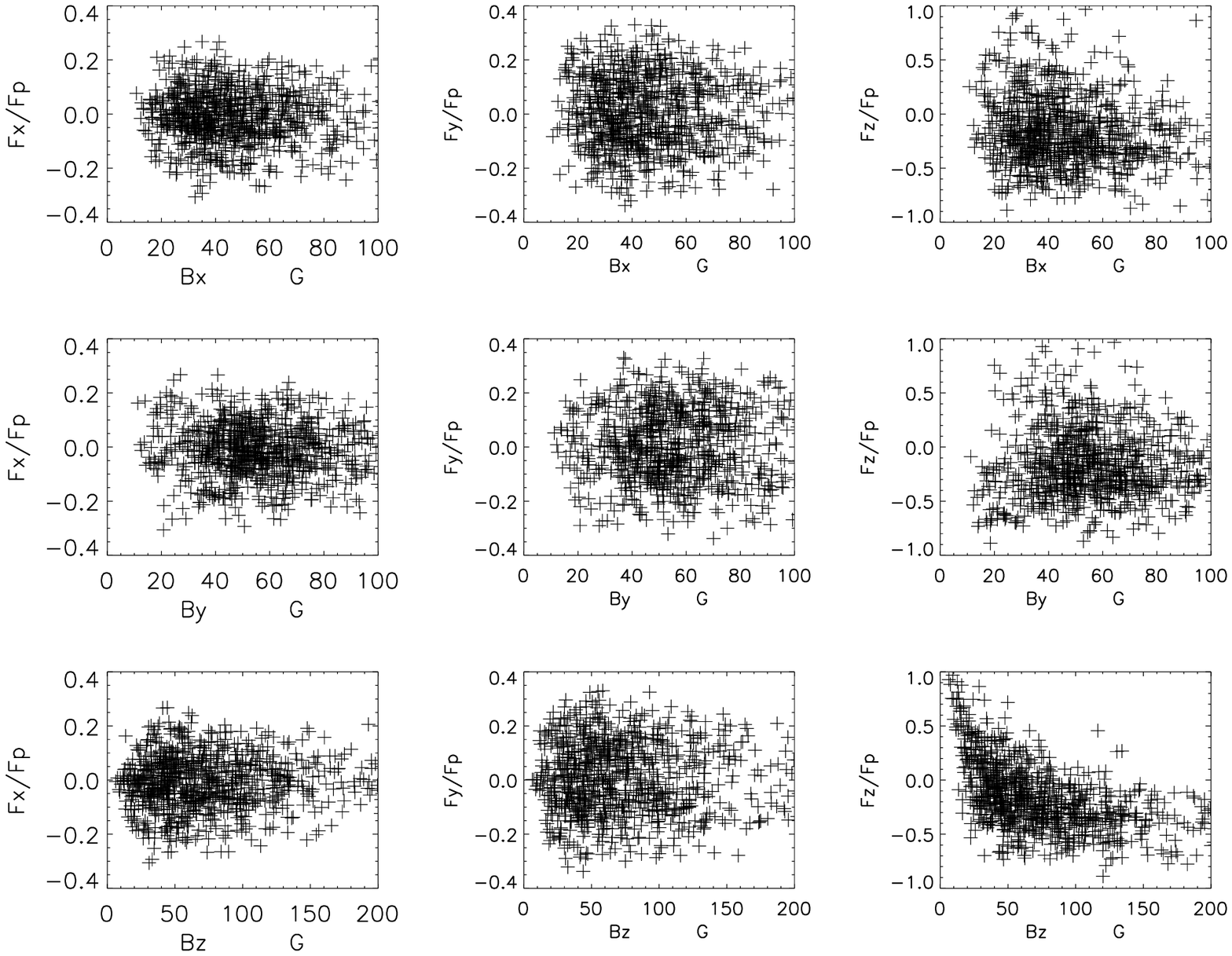}}

   \caption{$F_{x}/F_{p}$, $F_{y}/F_{p}$ and $F_{z}/F_{p}$ vs $B_{x}$, $B_{y}$ and $B_{z}$ for the selected magnetograms.
   (CaseI)} \label{Fig2}
\end{figure}

Fig \ref{Fig1} shows the possibility density function (PDF) and scatter
diagrams of $F_{x}/F_{p}$, $F_{y}/F_{p}$ and $F_{z}/F_{p}$ for the
selected magnetograms (CaseI). The mean values of absolute $F_{x}/F_{p}$,
$F_{y}/F_{p}$ and $F_{z}/F_{p}$ for all selected 925 magnetograms are 0.077,
0.109 and 0.302, respectively. The amplitudes of $F_{x}/F_{p}$ and $F_{y}/F_{p}$ are evidently smaller
than that of $F_{z}/F_{p}$, which is the same as previous study \citep{mat95}. Therefore $F_{z}/F_{p}$ can work as
a criterion indicating a force-free or non-force-free field more evidently. From the distributions of PDF
and scatter diagrams, the most of magnetograms
have the amplitudes of $F_{z}/F_{p}$ distributed outside the zone consisting the width of $\pm$~0.1, which means the most of the photospheric magnetic fields deviate from a force-free field. There are
about 17\% of the magnetograms with the values of $F_{z}/F_{p}$ less
than 0.1 for CaseI (38\% of the magnetograms with the values of $F_{z}/F_{p}$ less
than 0.2). To see the relation between $F_{x}/F_{p}$,
$F_{y}/F_{p}$ and $F_{z}/F_{p}$ and magnetic field strength, Fig
\ref{Fig2} shows $F_{x}/F_{p}$, $F_{y}/F_{p}$ and $F_{z}/F_{p}$ vs
the magnetic components of $B_{x}$, $B_{y}$ and $B_{z}$ for the selected
magnetograms (CaseI), where $B_{x}$, $B_{y}$ and $B_{z}$ are the
average of the absolute values of all pixels for each magnetogram.
It can be seen only at the bottom right of Fig \ref{Fig2} that the amplitudes of
$F_{z}/F_{p}$ decrease roughly as $B_{z}$ increase, and there is no
evident correlations exist between the parameters ($F_{x}/F_{p}$,
$F_{y}/F_{p}$ or $F_{z}/F_{p}$) and magnetic components, but according to Equation (\ref{fxyzp03}),
the value of $F_{z}/F_{p}$ should decrease as $B_{z}$ increasing.

To study the effect of calibration coefficients on these three
parameters of $F_{x}/F_{p}$, $F_{y}/F_{p}$ and $F_{z}/F_{p}$, PDF
and scatter diagrams of $F_{x}/F_{p}$, $F_{y}/F_{p}$ and
$F_{z}/F_{p}$ of the selected magnetograms (CaseII) are plotted in
Fig \ref{Fig3}. the amplitudes of $F_{x}/F_{p}$ and
$F_{y}/F_{p}$ same as CaseI are smaller than that of $F_{z}/F_{p}$ evidently. For
CaseII, The mean values of absolute $F_{x}/F_{p}$, $F_{y}/F_{p}$ and
$F_{z}/F_{p}$ for all selected magnetograms are 0.078, 0.111 and
0.251, respectively, which are smaller than those of CaseI
on the whole. And also for CaseII there are
about 25\% of the magnetograms with the value of $F_{z}/F_{p}$ less
than 0.1 (49\% of the magnetograms with the values of $F_{z}/F_{p}$ less
than 0.2). It should be noted that even more $F_{z}/F_{p}$ in the zone mentioned above (consisting the width of $\pm$~0.1), most
of magnetograms can not be regarded as a force-free.
Besides, the widths of PDF are not narrow and the scatter of
diagrams are diverge as well. In general, there is deviation of $F_{z}/F_{p}$ from the zone.
For CaseII, $F_{x}/F_{p}$, $F_{y}/F_{p}$, $F_{z}/F_{p}$ and magnetic components of
vs $B_{x}$, $B_{y}$ and $B_{z}$ are also plotted in Fig
\ref{Fig4} to study their relations with the magnetic field
strength. The results is consistent with that of CaseI,
only the the amplitudes of $F_{z}/F_{p}$ decrease as $B_{z}$
increasing.

To see the differences of results of $F_{x}/F_{p}$,
$F_{y}/F_{p}$ and $F_{z}/F_{p}$ between two cases, the scatter plots of
$F_{x}/F_{p}$, $F_{y}/F_{p}$ and $F_{z}/F_{p}$ of CaseI vs the corresponding ones from CaseII are shown in Fig \ref{Fig3}.
The correlations are calculated, which are
0.994, 0.995 and 0.980 for $F_{x}/F_{p}$, $F_{y}/F_{p}$ and $F_{z}/F_{p}$, respectively.
Also the linear fits ($y=Ax+B$) are done based on these scatter plots,
the values of A are 0.0008, 0.0003 and 0.2034 and B 1.10, 1.10 and 0.95 for
$F_{x}/F_{p}$, $F_{y}/F_{p}$ and $F_{z}/F_{p}$, respectively.
The correlation show there exists high consistence between these two cases.
Nevertheless correlation of $F_{z}/F_{p}$ between two cases is not so good as those of $F_{x}/F_{p}$ and
$F_{y}/F_{p}$. This may imply that the more attentions should be focused on the amplitude of $F_{z}/F_{p}$
in order to understand the extent of force-free.

\section{Discussions and Conclusions }
\label{S-Conl}

It is worth to study the force-free extent of the photospheric magnetic field while together with the case of magnetic field extrapolation,
since it has been assumed that the coronal magnetic field is force-free and the
photospheric one should be matched observationally. In this paper
the results of force-free extent of the photospheric magnetic field are
given through a statistical research, using 925 magnetograms corresponding to 925 active
regions observed by SMFT at HSOS.

A part of efforts to avoid the uncertainty of the data calibration is the employment
of two sets of calibration coefficients to describe
the force-free extent of the photospheric magnetic field. For the caseI, the calibration coefficients $C_{L}$ and
$C_{T}$ are 8381 G and 6790 G, the mean values of absolute $F_{x}/F_{p}$,
$F_{y}/F_{p}$ and $F_{z}/F_{p}$ for all selected magnetograms are
0.077, 0.109 and 0.302, respectively, and  for the caseII $C_{L}$ and $C_{T}$ are 10000 G and 9730 G, the mean values of absolute
$F_{x}/F_{p}$, $F_{y}/F_{p}$ and $F_{z}/F_{p}$ are 0.078, 0.111 and 0.251, respectively. There are 17\%
and 25\% magnetograms with their value of $F_{z}/F_{p}$ less than
0.1 for caseI and caseII, respectively, in other word, 17\% of caseI and 25\% of caseII are force-free. Although there are differences
between two cases, between them the correlation are as high as 0.994, 0.995 and 0.980,
for $F_{x}/F_{p}$, $F_{y}/F_{p}$ and $F_{z}/F_{p}$, respectively. Consequently we concluded that
large part of the photospheric magnetic fields do not belong to a force-free.
So before extrapolating a magnetic field, the
force-free extent of the photospheric magnetic field should be adequately considered.
We note that $F_{z}/F_{p}$ decrease with the increase of $B_{z}$, which is a more
important parameter indicating whether a magnetic
field is force-free or not, since the amplitude of $F_{z}/F_{p}$ is larger than those of
$F_{x}/F_{p}$ and $F_{y}/F_{p}$.
From Equation (\ref{fxyzp03}), it can be seen that $F_{z}/F_{p}$ may be neglected
when the amplitude of $B_{z}$ is comparable to $\sqrt{B_{x}^{2}+B_{y}^{2}}$,
while they are apparently different between $B_{z}$
and $\sqrt{B_{x}^{2}+B_{y}^{2}}$, then the amplitude of $F_{z}/F_{p}$ should be enlarged.
According to the Equations (\ref{fxyzp01}-(\ref{fxyzp03}) and observatories as well, the correlations between
$F_{x}/F_{p}$, $F_{y}/F_{p}$ and magnetic field component are not evident.

The strengths of active region magnetic fields observed by SMFT at HSOS
are determined through calibration, under the
conditions of a weak-field approximation and the linear relations
between the magnetic field and the Stokes parameters $I$, $Q$,
$U$ and $V$. Additionally, comparing the recent data obtained from
space satellite, the data of SMFT at HSOS has lower resolution
and more uncertainty of magnetic amplitudes. These disadvantages
may affect the statistical results we have acquired. However the
statistical analysis associated with its results may have an significance
as a practical reference, because SMFT has observed photospheric
magnetic fields more than one solar cycles at HSOS and its
reliability observations been studied strictly
and adequately. It is hoped that the statistical
results can be obtained based on high resolution data in the future.

\acknowledgments

The authors wish to thank the anonymous referee for his/her helpful
comments and suggestions.
This work was partly supported by the
 National Natural Science Foundation of China (Grant Nos. 10611120338, 10673016, 10733020,
10778723, 11003025, 11103037, 10878016 and 11178016), National Basic
Research Program of China (Grant No. 2011CB8114001), Important
Directional Projects of Chinese Academy of Sciences (Grant No.
KLCX2-YW-T04), the Young Researcher Grant of National
Astronomical Observatories, Chinese Academy of Sciences,
and Key aboratory of Solar Activity National Astronomical Observations, Chinese Academy of Sciences.


\begin{figure}

   \centerline{\includegraphics[width=1.\textwidth,clip=]{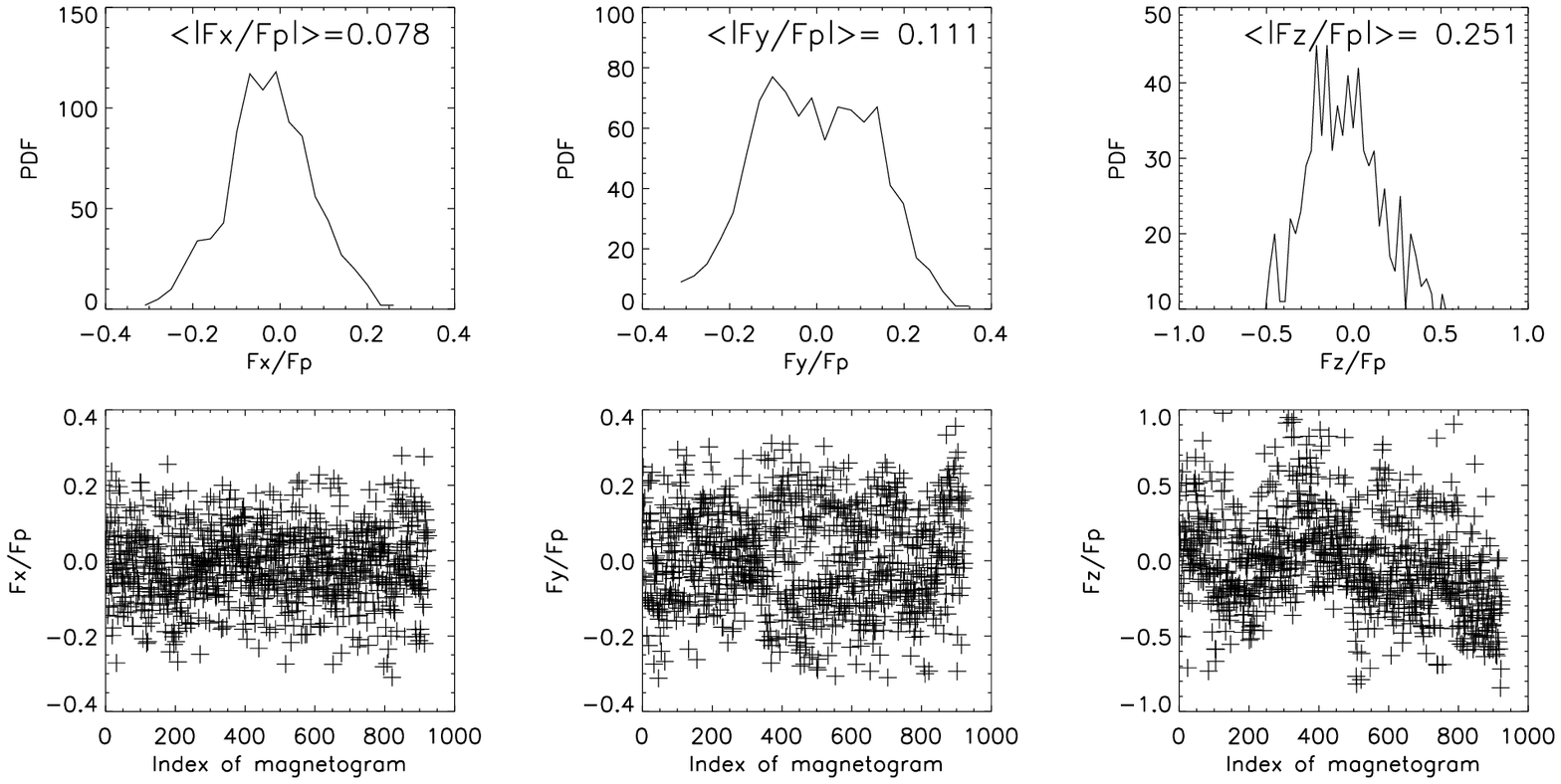}}

   \caption{PDF and scatter diagrams of $F_{x}/F_{p}$, $F_{y}/F_{p}$ and $F_{z}/F_{p}$ for the selected
   magnetograms. The mean values of absolute $F_{x}/F_{p}$, $F_{y}/F_{p}$ and
   $F_{z}/F_{p}$ are plotted and indicated by $<|F_{x}/F_{p}|>$,$<|F_{y}/F_{p}|>$, $<|F_{z}/F_{p}|>$, respectively. (CaseII)} \label{Fig3}
\end{figure}

\begin{figure}

   \centerline{\includegraphics[width=1.\textwidth,clip=]{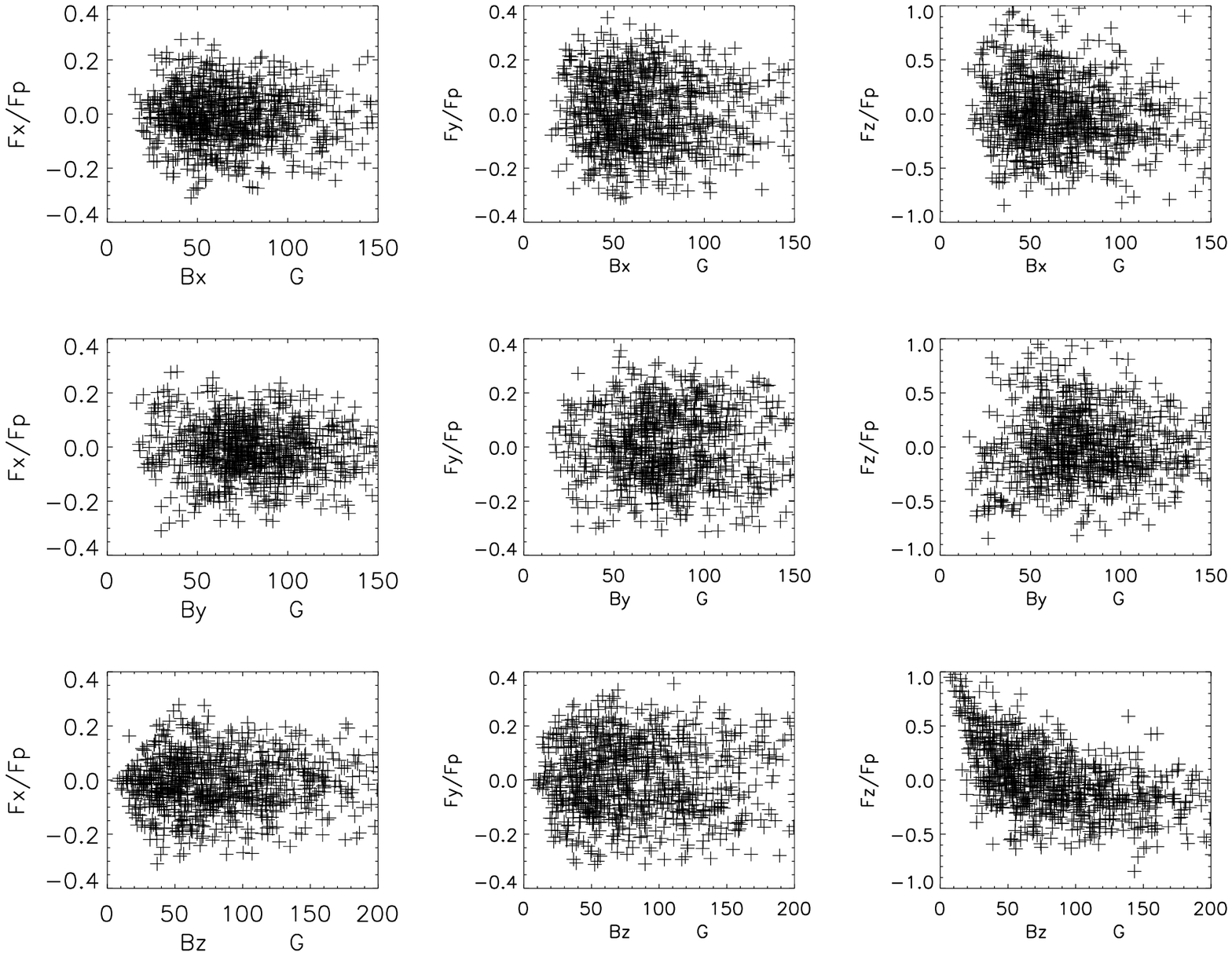}}

   \caption{$F_{x}/F_{p}$, $F_{y}/F_{p}$ and $F_{z}/F_{p}$ vs $B_{x}$, $B_{y}$ and $B_{z}$ for the selected magnetograms.
   (CaseII)} \label{Fig4}
\end{figure}

\begin{figure}

   \centerline{\includegraphics[width=1.\textwidth,clip=]{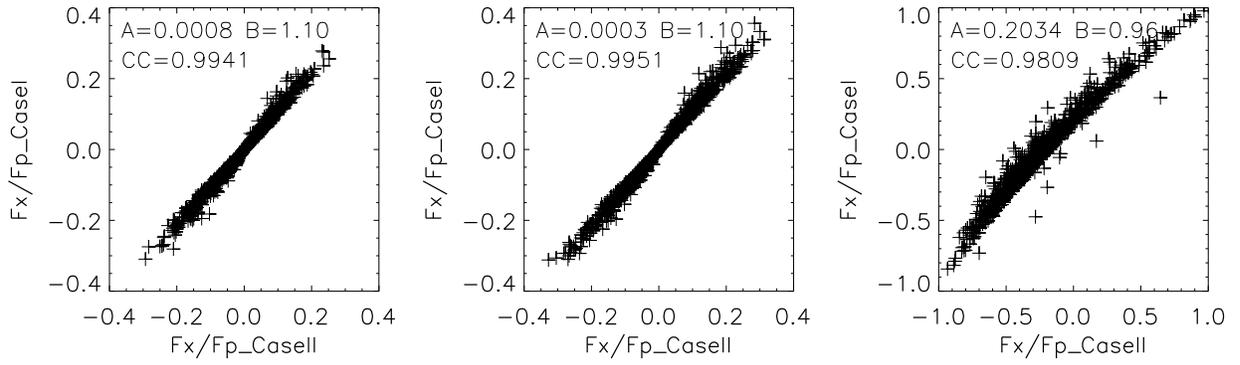}}

   \caption{Scatter diagrams of $F_{x}/F_{p}$, $F_{y}/F_{p}$ and $F_{z}/F_{p}$ deduced from CaseI and CaseII, respectively, for the selected
   magnetograms. } \label{Fig3}
\end{figure}





\end{document}